\documentclass[english]{article}
\usepackage[T1]{fontenc}
\usepackage[latin9]{inputenc}
\usepackage{geometry}
\usepackage{color}
\usepackage{array}
\usepackage{float}
\usepackage{amsmath}

\makeatletter

\providecommand{\tabularnewline}{\\}

\makeatother

\usepackage{babel}
\begin{document}

\title{A General Method for Selecting Quantum Channel for Bidirectional
Controlled State Teleportation and Other Schemes of Controlled Quantum
Communication}

\author{Kishore Thapliyal$^{a}$, Amit Verma$^{b}$ and Anirban Pathak$^{a,}$%
\thanks{anirban.pathak@jiit.ac.in%
}}

\maketitle
\begin{center}
$^{a}$Jaypee Institute of Information Technology, A-10, Sector-62,
Noida, UP-201307, India\\
$^{b}$Jaypee Institute of Information Technology, Sector-128, Noida,
UP-201304, India
\par\end{center}
\begin{abstract}
Recently, a large number of protocols for bidirectional controlled
state teleportation (BCST) have been proposed using $n$-qubit entangled
states ($n\in\{5,6,7\}$) as quantum channel. Here, we propose a general
method of selecting multi-qubit $(n>4)$ quantum channels suitable
for BCST and show that all the channels used in the existing protocols
of BCST can be obtained using the proposed method. Further, it is
shown that the quantum channels used in the existing protocols of
BCST forms only a negligibly small subset of the set of all the quantum
channels that can be constructed using the proposed method to implement
BCST. It is also noted that all these quantum channels are also suitable
for controlled bidirectional remote state preparation (CBRSP). Following
the same logic, methods for selecting quantum channels for other controlled
quantum communication tasks, such as controlled bidirectional joint
remote state preparation (CJBRSP) and controlled quantum dialogue,
are also provided. 
\end{abstract}

\section{Introduction\label{sec:Introduction-1}}

The idea of quantum teleportation was introduced by Bennett \textit{et
al}. \cite{Bennett} in 1993. Since this pioneering work, a large
number of modified teleportation schemes (such as, schemes for quantum
information splitting (QIS) or controlled teleportation (CT) \cite{Ct,A.Pathak},
quantum secret sharing (QSS) \cite{Hillery}, hierarchical quantum
information splitting (HQIS) \cite{hierarchical,Shukla}, remote state
preparation \cite{Pati}, etc.) have been proposed (see \cite{my-book}
for a review). Teleportation and its modified versions drew considerable
attention of the quantum communication community because of two main
reasons: firstly, teleportation is a purely quantum phenomenon having
no classical analogue and secondly, teleportation and modified teleportation
schemes have potential applications in secure quantum communication
and remote quantum operations \cite{J. A. Vaccaro}. 

Bennett \textit{et al}.'s original teleportation scheme \cite{Bennett}
enables the sender (Alice) to transmit an unknown single qubit quantum
state to the receiver (Bob) by using two bits of classical communication
and a pre-shared Bell state. This unidirectional scheme was subsequently
generalized by Huelga \textit{et al}. \cite{J. A. Vaccaro,J. A. Vaccaro-1}
and others by introducing protocols for bidirectional quantum state
teleportation (BST) which permit both Alice and Bob to simultaneously
transmit unknown quantum states to each other. Interestingly, Huelga
\textit{et al}. had also shown that nonlocal quantum gates can be
implemented using BST. Actually, the existence of a BST scheme ensures
the existence of a nonlocal quantum gate or a quantum remote control
(for a clearer discussion see our earlier works \cite{CBRSP-our-paper,bi-directional-ourpaper}).
This specific feature of BST attributed much importance to the study
of BST in context of both quantum computation and quantum communication.
In the recent past, the idea of BST has been further extended, and
a few schemes for bidirectional controlled state teleportation (BCST)
have been proposed \cite{bi-directional-ourpaper,Zha,Zha II,Li,5-qubit-c-qsdc,sixqubit1,six-qubit-3,six-qubit-4,7qubit}.
A standard BCST scheme is a three party scheme, where implementation
of BST is possible iff the controller (Charlie) allows the other two
users (Alice and Bob) to execute a protocol of BST \cite{bi-directional-ourpaper}.
A careful review of all the recently proposed BCST schemes \cite{Zha,Zha II,Li,5-qubit-c-qsdc,sixqubit1,six-qubit-3,six-qubit-4,7qubit}
that do not use permutation of particles (PoP) \cite{PoP,With chitra IJQI,With preeti}
technique reveals that different $n$-qubit (with $n\geq5)$ entangled
states are used in these protocols. In contrast, the same task can
also be achieved using Bell states and PoP as shown by some of the
present authors in Ref. \cite{crypt-switch}.\textcolor{blue}{{} }Clearly,
PoP-based schemes for BCST require lesser quantum resource as that
require only Bell states, whereas all the existing non-PoP based schemes
for BCST require at least 5 qubit entangled states. However, no prescription
for experimental realization of PoP exists until now. Keeping this
in mind, in the present paper, we restrict ourselves to non-PoP based
schemes for BCST. In an earlier work \cite{bi-directional-ourpaper},
some of the present authors explored the intrinsic symmetry of the
5-qubit quantum states that were used to propose the protocols of
BCST until then. In Ref. \cite{bi-directional-ourpaper}, following
general structure of the quantum states that can be used for BCST
was provided: 
\begin{equation}
|\psi\rangle=\frac{1}{\sqrt{2}}\left(|\psi_{1}\rangle_{A_{1}B_{1}}|\psi_{2}\rangle_{A_{2}B_{2}}|a\rangle_{C_{1}}\pm|\psi_{3}\rangle_{A_{1}B_{1}}|\psi_{4}\rangle_{A_{2}B_{2}}|b\rangle_{C_{1}}\right),\label{eq:the 5-qubit state-1}
\end{equation}
where single qubit states $|a\rangle$ and $|b\rangle$ satisfy $\langle a|b\rangle=\delta_{a,b}$,
$|\psi_{i}\rangle\in\left\{ |\psi^{+}\rangle,|\psi^{-}\rangle,|\phi^{+}\rangle,|\phi^{-}\rangle:|\psi_{1}\rangle\neq|\psi_{3}\rangle,|\psi_{2}\rangle\neq|\psi_{4}\rangle\right\} $,
$|\psi^{\pm}\rangle=\frac{|00\rangle\pm|11\rangle}{\sqrt{2}},$ $|\phi^{\pm}\rangle=\frac{|01\rangle\pm|10\rangle}{\sqrt{2}}$,
and the subscripts $A$, $B$ and $C$ indicate the qubits of Alice,
Bob and Charlie, respectively.\textcolor{blue}{{} }To illustrate that
$|\psi\rangle$ can be used to implement a BCST scheme, we may consider
that Charlie prepares the state $|\psi\rangle$ and sends 1st and
3rd (2nd and 4th) qubits to Alice (Bob) and keeps the 5th qubit with
himself. The condition 
\begin{equation}
|\psi_{1}\rangle\neq|\psi_{3}\rangle,|\psi_{2}\rangle\neq|\psi_{4}\rangle\label{eq:condition-1}
\end{equation}
ensures that Charlie's qubit is properly entangled with the remaining
4 qubits, and this in turn ensures that Alice and Bob are not aware
of the entangled (Bell) states they share until Charlie measures his
qubit using $\{|a\rangle,|b\rangle\}$ basis and announces the outcome
of his measurement. After Charlie's disclosure of the measurement
outcome, Alice and Bob know with certainty which two Bell states they
share, and that knowledge empowers them to use the conventional teleportation
scheme to teleport unknown quantum states to each other by using classical
communication and the unitary operations described in Table \ref{tab:table1}. 

\begin{table}
\begin{centering}
\begin{tabular}{|c|>{\centering}p{2cm}|>{\centering}p{2cm}|>{\centering}p{2cm}|>{\centering}p{2cm}|}
\hline 
 & \multicolumn{4}{c|}{Initial state shared by Alice and Bob}\tabularnewline
\cline{2-5} 
 & $|\psi^{+}\rangle$  & $|\psi^{-}\rangle$  & $|\phi^{+}\rangle$  & $|\phi^{-}\rangle$\tabularnewline
\cline{2-5} 
SMO  & Receiver's operation & Receiver's operation & Receiver's operation & Receiver's operation\tabularnewline
\hline 
00  & $I$  & $Z$  & $X$  & $iY$\tabularnewline
\hline 
01  & $X$  & $iY$  & $I$  & $Z$\tabularnewline
\hline 
10  & $Z$  & $I$  & $iY$  & $X$\tabularnewline
\hline 
11  & $iY$  & $X$  & $Z$  & $I$\tabularnewline
\hline 
\end{tabular}
\par\end{centering}

\caption{\label{tab:table1} The relation between the entangled states shared
by the receiver and sender, the measurement outcome of the sender,
and the unitary operations to be applied by the receiver to realize
perfect teleportation. Here SMO stands for the sender's measurement
outcome.}
\end{table}

In the above scheme for BCST, we require at least 5-qubit entanglement
and in that case we have to choose $|a\rangle$ and $|b\rangle$ as
single qubit states. However, we are allowed to take $|a\rangle$
and $|b\rangle$ as multi-qubit states and that leads to 6 or more
qubit quantum states capable of performing BCST. For example, in Refs.
\cite{sixqubit1,six-qubit-3,six-qubit-4}, BCST is reported using
6-qubit entangled states, and in Ref. \cite{7qubit}, BCST is reported
using a 7-qubit entangled state. These recent papers \cite{sixqubit1,six-qubit-3,six-qubit-4,7qubit}
on $n$-qubit ($n>5)$ implementation of BCST scheme, and the fact
that not all of them can be expressed in the form (\ref{eq:the 5-qubit state-1})
motivated us to extend our earlier work and to look for a general
structure of the $n$-qubit ($n\geq5)$ entangled states that can
be used to implement BCST. Keeping this in mind, here we construct
a general method for selecting quantum states for implementation of
multiqubit BCST schemes and subsequently extend the method to construct
suitable quantum states for other quantum communications tasks, such
as controlled bidirectional remote state preparation (CBRSP), controlled
joint bidirectional remote state preparation (CJBRSP), controlled
quantum dialogue, etc. In brief, we establish that there is a general
structure of the quantum states that are suitable for controlled quantum
communication tasks and there is not much merit in investigating specific
quantum channels in isolation.

Remaining part of the present paper is organized as follows. A general
method for selecting a quantum channel for multiqubit BCST is introduced
in Section \ref{sec:The-Condition-of}. In Section \ref{sec:Existing-states},
all the existing states used so far for BCST schemes are shown as
the special cases of the general structure introduced in the previous
section. In Section \ref{sec:cond-other-tasks}, methods for selecting
quantum channels for different controlled quantum communication tasks
(such as RSP, CBRSP, CJBRSP, controlled quantum dialogue, etc.) are
discussed. Finally, the paper is concluded in Section \ref{sec:Conclusions}.

\section{A general method for selecting a quantum channel for BCST\label{sec:The-Condition-of}}

The general structure of 5-qubit states used for BCST can further
be extended to the case where $|a\rangle$ and $|b\rangle$ in Eq.
(\ref{eq:the 5-qubit state-1}) are multiqubit states, and we can
write the general structure as
\begin{equation}
|\psi\rangle=\sum_{m=1}^{n}\frac{1}{\sqrt{n}}\left(\left(|\psi_{i}\rangle|\psi_{j}\rangle\right)_{m}|a_{m}\rangle\right),\label{eq:the multi-qubit state}
\end{equation}
where $|a_{m}\rangle$ are $n$ mutually orthogonal $l$-qubit states
with $2^{l}\geq n\geq2$, and $|\psi_{i}\rangle$ and $|\psi_{j}\rangle$
are the elements of a multiqubit basis set whose elements are maximally
(nonmaximally) entangled states capable of performing perfect (probabilistic)
teleportation (such as the set of Bell states or GHZ states) and $\left(|\psi_{i}\rangle|\psi_{j}\rangle\right)_{m}=\left(|\psi_{i}\rangle|\psi_{j}\rangle\right)_{m^{\prime}}{\rm \, iff}\, m=m^{\prime}$.
Further, if $|\psi_{i}\rangle$ and $|\psi_{j}\rangle$ are $p$-qubit
entangled states, then we can easily observe that $n\leq\left(2^{p}\right)^{2}.$
Thus, for the Bell states $n\leq16$, which implies that $l>4$ or
more than 8-qubit quantum states $|\psi\rangle$ used for the implementation
of BCST scheme involving Bell states for transmission of unknown qubits
in both the direction will not yield any new information (will not
represent a new quantum state in the perspective of information theory). 

Now, the states $|\psi_{i}\rangle$ and $|\psi_{j}\rangle$ are chosen
in such a way that unless Charlie measures his qubits in $\{|a_{m}\rangle\}$
basis and discloses the measurement outcome, Alice and Bob (in general
the receivers and the senders) do not know which entangled states
they share. This control of Charlie should exist in both the directions
of communication (i.e., prior to Charlie's disclosure the receiver
and the sender neither know the entangled state to be used for Alice
to Bob teleportation, nor they know the entangled state to be used
for Bob to Alice communication). A systematic method for obtaining
quantum states suitable for BCST may be described using the following
steps:
\begin{description}
\item [{Step~1:}] As $|\psi_{i}\rangle$ and $|\psi_{j}\rangle$ are the
elements of a basis set whose elements are $p$-qubit entangled states
we describe the basis set as $\left\{ |\psi_{i}\rangle:i\in\{1,2,\cdots,2^{p}\}\right\} $
and use that to construct a $2^{p}\times2^{p}$ matrix $S$ such that
$s_{ij}=|\psi_{i}\rangle|\psi_{j}\rangle$ is the $i^{th}$ row $j^{th}$
column element of the matrix 
\begin{equation}
S\equiv\left[\begin{array}{cccc}
|\psi_{1}\rangle|\psi_{1}\rangle & |\psi_{1}\rangle|\psi_{2}\rangle & \cdots & |\psi_{1}\rangle|\psi_{2^{p}}\rangle\\
|\psi_{2}\rangle|\psi_{1}\rangle & |\psi_{2}\rangle|\psi_{2}\rangle & \cdots & |\psi_{2}\rangle|\psi_{2^{p}}\rangle\\
\vdots & \vdots & \ddots & \vdots\\
|\psi_{2^{p}}\rangle|\psi_{1}\rangle & |\psi_{2^{p}}\rangle|\psi_{2}\rangle & \cdots & |\psi_{2^{p}}\rangle|\psi_{2^{p}}\rangle
\end{array}\right].\label{eq:Matrix S}
\end{equation}

\item [{Step~2:}] To construct a quantum state of the form (\ref{eq:the multi-qubit state})
that can perform BCST we choose $n\geq2$ elements of $S$ as $\left(|\psi_{i}\rangle|\psi_{j}\rangle\right)_{m}$
with the following restrictions.

\begin{description}
\item [{Rule~1:}] We cannot pick all the $n$ elements from the same row
or the same column of the matrix%
\footnote{If we choose all the elements from the $i^{th}$ row ($j^{th}$ column)
of $S$, then the desired quantum channel of the form (\ref{eq:the multi-qubit state})
will become separable, and Charlie will loose control in one direction
of the BST. Thus, the scheme will not remain BCST. %
} (\ref{eq:Matrix S}). 
\item [{Rule~2:}] We cannot pick one element more than once%
\footnote{This condition ensures the required bijective mapping between Charlie's
measurement outcome and the entangled states shared by Alice and Bob.
In the absence of this unique mapping, the receivers will not be able
to decide which unitary operation is to be applied to achieve teleportation.%
} as $\left(|\psi_{i}\rangle|\psi_{j}\rangle\right)_{m}=\left(|\psi_{i}\rangle|\psi_{j}\rangle\right)_{m^{\prime}}{\rm \, iff}\, m=m^{\prime}$.
\end{description}
\end{description}
Let us now elaborate the method described above using a simple example.
Consider that $\left\{ |\psi_{i}\rangle\right\} $ is a set of Bell
states and $|\psi^{+}\rangle=|\psi_{1}\rangle,\,|\psi^{-}\rangle=|\psi_{2}\rangle,\,|\phi^{+}\rangle=|\psi_{3}\rangle,\,{\rm and}\,|\phi^{-}\rangle=|\psi_{4}\rangle.$
Thus, the matrix (\ref{eq:Matrix S}) reduces to

\begin{equation}
S_{{\rm Bell}}\equiv\left[\begin{array}{cccc}
|\psi^{+}\rangle|\psi^{+}\rangle & |\psi^{+}\rangle|\psi^{-}\rangle & |\psi^{+}\rangle|\phi^{+}\rangle & |\psi^{+}\rangle|\phi^{-}\rangle\\
|\psi^{-}\rangle|\psi^{+}\rangle & |\psi^{-}\rangle|\psi^{-}\rangle & |\psi^{-}\rangle|\phi^{+}\rangle & |\psi^{-}\rangle|\phi^{-}\rangle\\
|\phi^{+}\rangle|\psi^{+}\rangle & |\phi^{+}\rangle|\psi^{-}\rangle & |\phi^{+}\rangle|\phi^{+}\rangle & |\phi^{+}\rangle|\phi^{-}\rangle\\
|\phi^{-}\rangle|\psi^{+}\rangle & |\phi^{-}\rangle|\psi^{-}\rangle & |\phi^{-}\rangle|\phi^{+}\rangle & |\phi^{-}\rangle|\phi^{-}\rangle
\end{array}\right].\label{eq:Choice-of-bell}
\end{equation}
 Now, consider that Charlie keeps a single qubit, and he measures
his qubit in $\left\{ |+\rangle,|-\rangle\right\} $ basis. Thus,
to construct a quantum state of the form (\ref{eq:the multi-qubit state})
we may consider $|a_{1}\rangle=|+\rangle$ and $|a_{2}\rangle=|-\rangle.$
Further, following the rules listed above, we may choose $\left(|\psi_{i}\rangle|\psi_{j}\rangle\right)_{1}=s_{{\rm Bell}_{11}}=|\psi^{+}\rangle|\psi^{+}\rangle$
and $\left(|\psi_{i}\rangle|\psi_{j}\rangle\right)_{2}=s_{{\rm Bell}_{22}}=|\psi^{-}\rangle|\psi^{-}\rangle$
as $s_{{\rm Bell}_{11}}\neq s_{{\rm Bell}_{22}},$ and they are not
elements of the same row or the same column. This choice would reduce
the quantum state described by (\ref{eq:the multi-qubit state}) to
$|\psi\rangle=\frac{1}{\sqrt{2}}\left(|\psi^{+}\rangle_{A_{1}B_{1}}|\psi^{+}\rangle_{A_{2}B_{2}}|+\rangle_{C_{1}}+|\psi^{-}\rangle_{A_{1}B_{1}}|\psi^{-}\rangle_{A_{2}B_{2}}|-\rangle_{C_{1}}\right)$.
This is the quantum channel used by Zha et al. in their proposal of
BCST \cite{Zha}.

Let us consider another example in which $n=4$ and $|a_{1}\rangle=|0+\rangle,\, a_{2}=|0-\rangle,\, a_{3}=|1+\rangle,\, a_{4}=-|1-\rangle$.
In order to construct a quantum state of the form (\ref{eq:the multi-qubit state}),
we have to select 4 elements of $S_{{\rm Bell}}$ in such a way that
Rules 1 and 2 are not violated. Keeping Rules 1 and 2 in mind, let
us select the elements of $S_{{\rm Bell}}$ shown in rectangular boxes
below as $\left(|\psi_{i}\rangle|\psi_{j}\rangle\right)_{m}$ 
\begin{equation}
S_{{\rm Bell}}\equiv\left[\begin{array}{cccc}
\boxed{|\psi^{+}\rangle|\psi^{+}\rangle} & \boxed{|\psi^{+}\rangle|\psi^{-}\rangle} & |\psi^{+}\rangle|\phi^{+}\rangle & |\psi^{+}\rangle|\phi^{-}\rangle\\
|\psi^{-}\rangle|\psi^{+}\rangle & |\psi^{-}\rangle|\psi^{-}\rangle & |\psi^{-}\rangle|\phi^{+}\rangle & |\psi^{-}\rangle|\phi^{-}\rangle\\
|\phi^{+}\rangle|\psi^{+}\rangle & |\phi^{+}\rangle|\psi^{-}\rangle & \boxed{|\phi^{+}\rangle|\phi^{+}\rangle} & \boxed{|\phi^{+}\rangle|\phi^{-}\rangle}\\
|\phi^{-}\rangle|\psi^{+}\rangle & |\phi^{-}\rangle|\psi^{-}\rangle & |\phi^{-}\rangle|\phi^{+}\rangle & |\phi^{-}\rangle|\phi^{-}\rangle
\end{array}\right].\label{eq:Ex.Choice-of-bell}
\end{equation}
Note that neither all the elements shown in rectangular boxes belong
to the same row nor they belong to the same column. Thus, they satisfy
the rules. Further, if we arrange the selected elements as $\left(|\psi_{i}\rangle|\psi_{j}\rangle\right)_{1}=|\psi^{+}\rangle|\psi^{+}\rangle,\,\left(|\psi_{i}\rangle|\psi_{j}\rangle\right)_{2}=|\psi^{+}\rangle|\psi^{-}\rangle$,
$\left(|\psi_{i}\rangle|\psi_{j}\rangle\right)_{3}=|\phi^{+}\rangle|\phi^{+}\rangle,\,\left(|\psi_{i}\rangle|\psi_{j}\rangle\right)_{4}=|\phi^{+}\rangle|\phi^{-}\rangle$,
then we obtain

\begin{equation}
\begin{array}{c}
|\psi\rangle=\frac{1}{2}\left(|\psi^{+}\rangle_{A_{1}B_{1}}|\psi^{+}\rangle_{A_{2}B_{2}}|0+\rangle_{C_{1}C_{2}}+|\psi^{+}\rangle_{A_{1}B_{1}}|\psi^{-}\rangle_{A_{2}B_{2}}|0-\rangle_{C_{1}C_{2}}\right.\\
\left.+|\phi^{+}\rangle_{A_{1}B_{1}}|\phi^{+}\rangle_{A_{2}B_{2}}|1+\rangle_{C_{1}C_{2}}-|\phi^{+}\rangle_{A_{1}B_{1}}|\phi^{-}\rangle_{A_{2}B_{2}}|1-\rangle_{C_{1}C_{2}}\right)
\end{array}\label{eq:example2}
\end{equation}
which is the quantum state used in Ref. \cite{six-qubit-3} to implement
BCST (cf. Eq. (1) of \cite{six-qubit-3}). Now, we may choose the
same $|a_{1}\rangle,|a_{2}\rangle,|a_{3}\rangle,$ and $|a_{4}\rangle$
as used in this example and select 4 other elements of $S_{{\rm Bell}}$
that are not from the same row/column to obtain a new 6-qubit quantum
channel (one possible quantum channel for each selection) that can
be used to realize BCST. Thus, we observe that if we follow the method
prescribed here, we can easily generate several new quantum states
that can be used to implement BCST. In what follows, we will show
that the number of possible quantum channels is extremely high and
only a few possibilities have been studied in the existing works on
BCST. 

In the first example above, Charlie keeps only one qubit with himself,
and consequently we were required to choose two elements of $S_{{\rm Bell}}$
without violating rules 1 and 2 stated above. A specific example is
shown above. However, the rules allow us to choose any of the 16 elements
of $S_{{\rm Bell}}$ as $\left(|\psi_{i}\rangle|\psi_{j}\rangle\right)_{1}.$
The moment we make a specific choice, Rule 1 allows us to choose $\left(|\psi_{i}\rangle|\psi_{j}\rangle\right)_{2}$
only from 9 elements of $S_{{\rm Bell}}$ (i.e., for the second choice,
the row and column of the first choice are exempted). Thus, the total
possible choices of 5-qubit quantum states that can implement BCST
is $16\times9=144$ for a specific choice of Charlie's measurement
basis $\left\{ |a_{m}\rangle\right\} .$ This coincides exactly with
the results reported in Ref. \cite{bi-directional-ourpaper}. Extending
this logic to more general cases, we may note that a simple algebraic
analysis reveals that for a specific choice of a subset of order $n$
of a basis set $\{|a_{m}\rangle\},$ the total number of possible
states $(N_{s})$ of the form (\ref{eq:the multi-qubit state}) that
can be constructed using $S$ without violating the rules mentioned
above is 
\begin{equation}
N_{s}=\left\{ \begin{array}{cc}
\frac{2^{2p}!}{\left(2^{2p}-n\right)!} & {\rm for}\, n>2^{p}\\
2^{pn}\left(2^{pn}-2^{p+1}+1\right) & {\rm for}\, n<2^{p}
\end{array}\right..\label{eq:NS}
\end{equation}
Using (\ref{eq:NS}), we may quickly obtain the possible number of
quantum channels for Bell-state based BCST for a specific choice of
\{$|a_{m}\rangle$\} using $p=2$. Specifically, in this case, if
$n>4$, then $N_{s}=\frac{16!}{\left(16-n\right)!}$ and consequently,
for $n=2,3$ and 4, we can construct quantum states of the form (\ref{eq:the multi-qubit state})
in 144, 3648 and 63744 ways, respectively. These numbers clearly show
that until now BCST is investigated using a very small subset of all
possible states that can be used to realize BCST. Here, we have obtained
a quantitative measure of $N_{s}$ for a specific choice of an $n$th
order subset of $\{|a_{m}\rangle\}$ without considering the following:
(i) the relative phases of the superposition in (\ref{eq:the multi-qubit state}),
(ii) possible permutations of $\{|a_{m}\rangle\}$ and (iii) the number
of ways in which subset of order $n$ can be constructed. Inclusion
of these factors will further enhance the number of ways in which
BCST can be done. Specifically, if Charlie keeps $l$ qubits with
himself, then the inclusion of the last two factors listed above will
further increase the total number of possible states by $\frac{2^{l}!}{\left(2^{l}-n\right)!}$
times for specific value of $n$. Our intention is not to obtain the
total number. We are interested to establish that there exists a systematic
way to obtain quantum states that can implement BCST, and the set
of all the states that are shown to be useful in implementing BCST
only forms a small subset of the set of all possible states that can
be used to implement BCST. Above discussion firmly establishes the
fact we intended to establish.

\section{Existing states as the special cases of the general structure\label{sec:Existing-states}}

In Section \ref{sec:Introduction-1}, we have already mentioned that
schemes for BCST have been proposed in the recent past using various
quantum states. Here, we show that all the states used till date can
be expressed in the general form (\ref{eq:the multi-qubit state}).
This fact is explicitly shown in Table \ref{tab:special-cases}, where
following notation is used to express GHZ states:\textcolor{red}{{} }

\begin{equation}
{\rm GHZ^{x\pm}}=\frac{\left(|iij\rangle\pm|\bar{i}\bar{i}\bar{j}\rangle\right)}{\sqrt{2}},\label{eq:vishal1}
\end{equation}
where $x$ is the decimal value of binary number $iij$ with $i,j\in\left\{ 0,1\right\} ,$
and $\pm$ denotes the relative phase between the two components of
the superposition. For example, if we consider $i=j=0$, we obtain
${\rm GHZ^{0\pm}}=\frac{\left(|000\rangle\pm|111\rangle\right)}{\sqrt{2}}$. 

\begin{table}[H]
\centering{}%
\begin{tabular}{|>{\centering}p{3cm}|>{\centering}p{11.5cm}|>{\centering}p{2cm}|}
\hline 
Quantum states used in existing works & How to express the quantum states used in the existing work in the
generalized form described in the present paper? & Remarks\tabularnewline
\hline 
Eq. (1) in Ref. \cite{Zha} & $\frac{1}{\sqrt{2}}\left(|\psi^{+}\rangle_{A_{1}B_{1}}|\psi^{+}\rangle_{A_{2}B_{2}}|+\rangle_{C_{1}}+|\psi^{-}\rangle_{A_{1}B_{1}}|\psi^{-}\rangle_{A_{2}B_{2}}|-\rangle_{C_{1}}\right)$  & \tabularnewline
\hline 
Eq. (8) in Ref. \cite{Zha II} & $\frac{1}{\sqrt{2}}\left(|\psi^{+}\rangle_{A_{1}B_{1}}|\psi^{+}\rangle_{A_{2}B_{2}}|0\rangle_{C_{1}}-|\psi^{-}\rangle_{A_{1}B_{1}}|\phi^{-}\rangle_{A_{2}B_{2}}|1\rangle_{C_{1}}\right)$ & \tabularnewline
\hline 
Eq. (3) in Ref. \cite{Li} & $\frac{1}{\sqrt{2}}\left(|\psi^{+}\rangle_{A_{1}B_{1}}|+\rangle_{C_{1}}+|\psi^{-}\rangle_{A_{1}B_{1}}|-\rangle_{C_{1}}\right)|\psi^{+}\rangle_{A_{2}B_{2}}$  & Charlie's control is limited to\tabularnewline
\cline{1-2} 
Eq. (3) in Ref. \cite{5-qubit-c-qsdc} & $\frac{1}{\sqrt{2}}\left(|\psi^{+}\rangle_{A_{2}B_{2}}|0\rangle_{C_{1}}+|\psi^{-}\rangle_{A_{2}B_{2}}|1\rangle_{C_{1}}\right)|\phi^{+}\rangle_{A_{1}B_{1}}$ & one side of teleportation.\tabularnewline
\hline 
Eq. (12) in Ref. \cite{sixqubit1} & $\frac{1}{\sqrt{2}}\left(|\psi^{+}\rangle_{A_{1}B_{1}}|\phi^{+}\rangle_{A_{2}B_{2}}|00\rangle_{C_{1}C_{2}}+|\phi^{+}\rangle_{A_{1}B_{1}}|\psi^{+}\rangle_{A_{2}B_{2}}|11\rangle_{C_{1}C_{2}}\right)$  & \tabularnewline
\hline 
Eq. (1) in Ref. \cite{six-qubit-3} & $\begin{array}{c}
\frac{1}{2}\left(|\psi^{+}\rangle_{A_{1}B_{1}}|\psi^{+}\rangle_{A_{2}B_{2}}|0+\rangle_{C_{1}C_{2}}+|\psi^{+}\rangle_{A_{1}B_{1}}|\psi^{-}\rangle_{A_{2}B_{2}}|0-\rangle_{C_{1}C_{2}}\right.\\
\left.+|\phi^{+}\rangle_{A_{1}B_{1}}|\phi^{+}\rangle_{A_{2}B_{2}}|1+\rangle_{C_{1}C_{2}}-|\phi^{+}\rangle_{A_{1}B_{1}}|\phi^{-}\rangle_{A_{2}B_{2}}|1-\rangle_{C_{1}C_{2}}\right)
\end{array}$ & \tabularnewline
\hline 
Eq. (3) in Ref. \cite{six-qubit-4} & $\begin{array}{c}
\frac{1}{2}\left(|\psi^{+}\rangle_{A_{1}B_{1}}|\psi^{+}\rangle_{A_{2}B_{2}}|++\rangle_{C_{1}C_{2}}+|\psi^{+}\rangle_{A_{1}B_{1}}|\phi^{-}\rangle_{A_{2}B_{2}}|+-\rangle_{C_{1}C_{2}}\right.\\
\left.+|\phi^{-}\rangle_{A_{1}B_{1}}|\psi^{+}\rangle_{A_{2}B_{2}}|-+\rangle_{C_{1}C_{2}}+|\phi^{-}\rangle_{A_{1}B_{1}}|\phi^{-}\rangle_{A_{2}B_{2}}|--\rangle_{C_{1}C_{2}}\right)
\end{array}$

or

$\begin{array}{c}
\frac{1}{2}\left(|\psi^{+}\rangle_{A_{1}B_{1}}|\phi^{+}\rangle_{A_{2}B_{2}}|0+\rangle_{C_{1}C_{2}}+|\psi^{+}\rangle_{A_{1}B_{1}}|\phi^{+}\rangle_{A_{2}B_{2}}|0-\rangle_{C_{1}C_{2}}\right.\\
\left.+|\phi^{+}\rangle_{A_{1}B_{1}}|\psi^{+}\rangle_{A_{2}B_{2}}|1+\rangle_{C_{1}C_{2}}-|\phi^{+}\rangle_{A_{1}B_{1}}|\psi^{+}\rangle_{A_{2}B_{2}}|1-\rangle_{C_{1}C_{2}}\right)
\end{array}$  & These are shown in Ref. \cite{sixqubit1} as Eqs. (11) and (13).\tabularnewline
\hline 
Eq. (4) in Ref. \cite{7qubit} & $\begin{array}{c}
\frac{1}{2}\left(|\psi^{+}\rangle_{A_{1}B_{1}}|\psi^{+}\rangle_{A_{2}B_{2}}|{\rm GHZ}^{0+}\rangle_{C_{1}C_{2}C_{3}}-|\psi^{-}\rangle_{A_{1}B_{1}}|\phi^{-}\rangle_{A_{2}B_{2}}|{\rm GHZ}^{2+}\rangle_{C_{1}C_{2}C_{3}}\right.\\
\left.-|\phi^{+}\rangle_{A_{1}B_{1}}|\phi^{+}\rangle_{A_{2}B_{2}}|{\rm GHZ}^{3+}\rangle_{C_{1}C_{2}C_{3}}-|\phi^{-}\rangle_{A_{1}B_{1}}|\psi^{-}\rangle_{A_{2}B_{2}}|{\rm GHZ}^{1+}\rangle_{C_{1}C_{2}C_{3}}\right)
\end{array}$ & \tabularnewline
\hline 
\end{tabular}\caption{\label{tab:special-cases}Quantum channels used in different proposals
for BCST as special cases of the generalized structure shown here.}
\end{table}

\section{The condition for selecting a quantum channel for other controlled
quantum communication tasks\label{sec:cond-other-tasks}}

Several schemes of controlled quantum communication have been discussed
in the recent past (\cite{CBRSP-our-paper,bi-directional-ourpaper,crypt-switch,ba-An-remote-state,C-QD1,C-QD2,CQSDC-Hassanpour,switch}
and references therein). To be precise, schemes for controlled bidirectional
remote state preparation \cite{CBRSP-our-paper,ba-An-remote-state},
controlled joint bidirectional remote state preparation \cite{CBRSP-our-paper},
controlled quantum dialogue \cite{C-QD1,C-QD2}, etc., have been proposed
using various quantum states. Extending the argument above, in what
follows, we provide a general method for selecting quantum states
for these tasks. Here, we limit ourselves to the explicit discussion
of the general structure for the quantum states required for (i) controlled
bidirectional remote state preparation, (ii) controlled joint bidirectional
remote state preparation and (iii) controlled quantum dialogue, but
the logic can be extended easily to other controlled quantum communication
tasks.

\subsection{How to select a\textcolor{magenta}{{} }quantum channel for Controlled
Bidirectional Remote State Preparation?}

In Ref. \cite{CBRSP-our-paper}, some of the present authors have
shown that the quantum states suitable for BCST are also suitable
for CBRSP. This is reasonable for the obvious reason that the capability
of transportation of an unknown state automatically implies the capability
of transporting a known quantum state. Further, it is well known that
a shared Bell state and one bit of classical communication is sufficient
for probabilistic RSP \cite{Pati}, whereas a shared Bell state and
two bits of classical communication is sufficient for deterministic
RSP. This fact and our discussion above in the context of the choice
of quantum states for controlled BCST imply that the quantum states
of the form (\ref{eq:the multi-qubit state}) are sufficient for CBRSP
if $|\psi_{i}\rangle,\,|\psi_{j}\rangle$ are chosen using the rules
described above.

\subsection{How to select a quantum channel for Controlled Joint Bidirectional
Remote State Preparation?}

For CJBRSP, the structure of quantum state to be used would remain
same (i.e., the states described by Eq. (\ref{eq:the multi-qubit state})
and element selection rules described after that) with the only difference
that $|\psi_{i}\rangle$ and $|\psi_{j}\rangle$ must be the elements
of a basis set whose elements are at least tripartite entangled and
capable of performing joint remote state preparation. Specifically,
$|\psi_{i}\rangle$ and $|\psi_{j}\rangle$ can be GHZ or GHZ-like
states.

\subsection{How to select a quantum channel for Controlled Quantum Dialogue?}

In case of quantum dialogue protocols of Ba-An type \cite{ba-an,qd},
the quantum communication happens in both directions using the same
quantum state, hence we do not require product of two entangled states
after the measurement of controller Charlie. Here, it is sufficient
to choose a quantum state such that unless the controller measures
his/her qubit and announces the outcome, other two users (Alice and
Bob) will be unaware of the quantum state they share. Thus, any quantum
state of the form 
\begin{equation}
|\psi\rangle=\sum_{m=1}^{n}\frac{1}{\sqrt{n}}\left(|\psi_{i}\rangle|a_{m}\rangle\right),\label{eq:the state for QD}
\end{equation}
where $|a_{m}\rangle$ are $n$ mutually orthogonal $l$-qubit states
with $2^{l}\geq n\geq2$, and $|\psi_{i}\rangle$ is an element of
a set of entangled quantum states that are capable of performing quantum
dialogue using the same set of unitary operators and that are unitarily
connected with each other. Such that after the encoding operation
of Alice (say $U_{j})$ and that of Bob (say $U_{i}$) the final states
must also be a member of the set of mutually orthogonal states to
ensure the deterministic discrimination of the state and thus to decode
the encoded message, where $U_{j}$ and $U_{i}$ are the unitary operators
which forms a group under multiplication. To be precise, $|\psi\rangle_{final}=U_{j}U_{i}|\phi_{0}\rangle=U_{j}|\phi_{i}\rangle\in\{|\phi_{0}\rangle,|\phi_{1}\rangle,\cdots,|\phi_{i}\rangle,\cdots,|\phi_{2^{n}-1}\rangle\,\forall\, i,j\in\{0,1,\cdots,2^{n}-1\}\Rightarrow U_{B}U_{A}\in\{U_{0},U_{1},U_{2},\cdots,U_{2^{n}-1}\}$.
A list of quantum states that can be used for quantum dialogue protocol
with corresponding unitary operators are given in Table 4 in Ref.
\cite{qd}. Superpositions of such states with mutually orthogonal
states in the control part $|a_{m}\rangle$ can be used for the generalized
protocol of controlled quantum dialogue.

\section{Conclusions\label{sec:Conclusions}}

The general structures of the quantum states suitable for BCST and
other controlled quantum communication protocols are provided, and
a method for obtaining all such states is proposed. Further, it is
shown that all the quantum channels used in the existing protocols
of BCST can be easily obtained using the general method proposed here.
In fact, all the quantum states that are used in the existing protocols
are explicitly expressed in the general form proposed here (cf. Table
\ref{tab:special-cases}). The states described in Table \ref{tab:special-cases},
i.e., the states used in the existing literature, only forms a negligibly
small subset of all possible states that can perform BCST. It is easy
to visualize that there are infinitely many possible states that can
be used to perform BCST. To elaborate this point we may note that
there is no constraint on the choice of the set of controllers multiqubit
orthogonal states. The infinitely many possible choices for the set
of controllers multiqubit orthogonal states imply availability of
infinitely many possible quantum channels for BCST. Even if we restrict
Charlie to prepare and measure his qubits in a specific basis (say,
computational basis) and Alice and Bob to use Bell states for quantum
communication, there exist a large number of ways in which quantum
states of the form (\ref{eq:the multi-qubit state}) can be constructed.
This point is firmly established in Section \ref{sec:The-Condition-of}. 

We have already seen that the general structure provided here for
BCST scheme gives an infinitely many possibilities for the choice
of the suitable quantum channels to experimentalists. The generation
of this kind of quantum states suitable for BCST scheme requires easily
available resources, such as CNOT gate, Hadamard gate, etc. A multiqubit
quantum states of the form given here (i.e., of the form (\ref{eq:the multi-qubit state}))
have already been experimentally realized in the recent past \cite{6-qubit-cat-nature},\textcolor{red}{{}
}and further discussion on the possibilities of experimental preparation
of the quantum states of the structure similar to the structure given
here can be found in our earlier work \cite{CBRSP-our-paper}. Further,
if we consider that $|\psi_{i}\rangle$ and $|\psi_{j}\rangle$ are
Bell states, then after the measurement of the controller's (Charlie's)
qubits in suitable basis, the remaining qubits reduce to a product
state $|\psi_{i}\rangle\otimes|\psi_{j}\rangle$, which is the product
of two entangled states (product of two Bell states in the case of
Bell state based BCST scheme) shared between Alice and Bob, and which
can be used for simultaneous teleportation of two unknown quantum
states, one from Alice to Bob and the other from Bob to Alice. The
resources required in BCST after the Charlie's measurement are just
two copies of the resources required for teleportation of the unknown
quantum state using the entangled state shared by Alice and Bob.

\textbf{Acknowledgment:} AP and KT thank Department of Science and
Technology (DST), India for support provided through the DST project
No. SR/S2/LOP-0012/2010.

\end{document}